\documentclass[aps,pra,showpacs,twocolumn,groupedaddress]{revtex4}
\usepackage{amsmath}
\usepackage{amssymb}
\usepackage{graphics}
\usepackage[dvips]{graphicx}
\usepackage{graphicx}% Include figure files
\usepackage{color}

\newcommand \beq{\begin{eqnarray}}
\newcommand \eeq{\end{eqnarray}}
\def\simge{\mathrel{%
       \rlap{\raise 0.511ex \hbox{$>$}}{\lower 0.511ex \hbox{$\sim$}}}}
\def\simle{\mathrel{
       \rlap{\raise 0.511ex \hbox{$<$}}{\lower 0.511ex \hbox{$\sim$}}}}

\begin{document}
\title{Hanbury Brown--Twiss interferometry with electrons:
  Coulomb vs. quantum statistics\footnote{Contribution to the {\em Memorial Publication in honor of Akira Tonomura}, 
[eds. K. Fujikawa and Y. A. Ono] to be published by World Scientific Publishing.}}
\author{Gordon Baym$^{a}$ and Kan Shen$^{a,b}$}
\affiliation{\mbox{$^a$Department of Physics, University of Illinois, 1110
  W. Green Street, Urbana, IL 61801} \\
\mbox{$^b$Quantitative Strategies, Credit Suisse, 11 Madison Ave, New York, NY 10010C } 
}

\date{\today}

\begin{abstract}

A longstanding goal of Akira Tonomura  was to observe
Hanbury Brown--Twiss anti-correlations between electrons in a field-emission free electron beam.  The experimental results were reported in his 2011 paper
with Tetsuji Kodama and
Nobuyuki Osakabe~\cite{KOT}.  An open issue in such a measurement is whether the observed anti-correlations arise
from quantum statistics, or are simply produced by Coulomb repulsion between
electrons.  In this paper we describe a simple classical model of Coulomb effects to
estimate their effects in electron beam interferometry experiments, and conclude that the experiment did indeed observe quantum correlations in the electron arrrival times.

\end{abstract}

\maketitle

\section{Introduction}

Since the pioneering detection by Hanbury Brown and Twiss (HBT) of ``bunching" of photons in a light beam  \cite{hanbury},  HBT experiments with massive bosons, e.g.,
atomic beams~\cite{shimizu, jeltes} and $\pi$ and $K$ mesons in
high energy nuclear collisions \cite{boal,zakopane} have shown similar
two-particle correlations.  Seeing anti-correlation -- or anti-bunching -- effects in experiments
with identical fermions where the two-particle intensity ($I$)
correlation function
\begin{equation}
  C(\mathbf{r}_2-\mathbf{r}_1,t_2-t_1)
  = \frac{\langle I_1 I_2\rangle }{\langle I_1\rangle \langle I_2\rangle}.
\end{equation}
should fall, at small separations (either in position or momentum
space), to zero for particles of the same spin (or to 1/2 for unpolarized particles) has proven more elusive.  Experiments with neutrons are complicated by
a low energy nuclear resonance, while experiments with protons are
complicated in addition by Coulomb repulsion~\cite{bauer, ghetti}.  
 On the other hand.
anti-bunching of
neutral cold fermionic $^{40}$K atoms \cite{rom}, and the corresponding
bunching of neutral cold bosonic $^{87}$Rb atoms \cite{folling} emitted from optical lattices has been successfully observed.
In adddition, anti-bunching
with neutral atomic $^3$He beams, as well as bunching with neutral atomic
$^4$He atomic beams, was clearly demonstrated in the experiment of Jeltes et al.~\cite{jeltes}.

Detecting anti-bunching in a beam of electrons has been a 
major experimental challenge over the years, owing to the low degeneracy
as well as the short coherence time of the beams.
Starting in the 1990's Akira Tonomura and his group focussed on
seeing this striking effect of quantum statistics with electrons in a field-emission electron beam.
Following his group's theoretical feasibility analysis \cite{tonomura1,
tonomura2}, electron HBT experiments have been realized in free space
\cite{kiesel,KOT};   such
experiments with electrons show a reduction in the correlation function for
small space-time separation, generally attributed to anti-bunching.  On the
other hand, repulsive Coulomb interactions between electrons also reduce the
probability of two electrons being close in space.  Whether the observed anti-bunching effect is due to
electron quantum statistics or rather Coulomb repulsion is the issue we deal
with in this paper.  We conclude that the recently reported experiment of Kodama, Osakabe, and Tonomura \cite{KOT}
very cleanly sees HBT in the arrival time correlations of electrons pairs;
In the experiment of  Ref.~14, at a significantly
lower beam energy, Coulomb effects account for several percent of the HBT signal.

   HBT interferometry has in fact been seen for electrons in semiconductor devices, \cite{henny, oliver} where, owing to screening, Coulomb effects are less important.  For example, in the HBT experiment of Ref. 16, 
the screening length ($\sim$ 5nm) is typically much smaller than the Fermi wavelength
($\sim$ 40nm) \cite{oliverTh}.   Interestingly, two dimensional mesoscopic semiconductors open the possibility of seeing
HBT correlations for fractional statistics  \cite{smitha,campagnolo} as well as Aharonov-Bohm physics \cite{smitha2}.

    The importance of correcting for Coulomb interactions has long been recognized in interpreting
high energy nuclear experiments \cite{pratt}.  For example, the raw correlation function of
distinguishable pions of opposite charge ($\pi^+\pi^-$), produced in the E877
ultra-relativistic heavy ion collision experiment \cite{E877}, shows a very
similar buildup at small momentum differences to those of identical charged
pions ($\pi^+\pi^+$ and $\pi^-\pi^-$).  The Coulomb interaction between
opposite charges tends to increase the probability of a pair of bosons being
close in momentum, while reducing that for like charges.  Only after the
effects of Coulomb interactions are extracted, does one see the expected
effects of quantum statistics (see, e.g., Refs. 7 and 22).

   Our aim in this note is to present a simple schematic discussion of Coulomb effects in interferometry experiments with electrons, based on the classical behavior of electrons taken pairwise.   We do 
not attempt to explain the detailed results of Tonomura's group on HBT with
electrons, but rather aim to estimate the role of Coulomb interactions in their search for
quantum correlations.

    The Coulomb problem for a pair of electrons is characterized by four
length scales, 1) the electron Bohr radius $a_0 \equiv \hbar^2/m
e^2$, with $m$ the electron mass; 2) the size $r_0$ of the emitting region transverse to the beam; 3) the typical separation  $z_0$ of the particles along the beam
direction; and importantly, 4) the classical turning point of
the pair, $r_{tp}$, defined by
$
 e^2/r_{tp} = q^2/2m_{red},
$
where $q$ is magnitude of the final relative momentum of the two
particles and $m_{red} = m/2$.

\begin{figure}
\includegraphics[width=5cm]{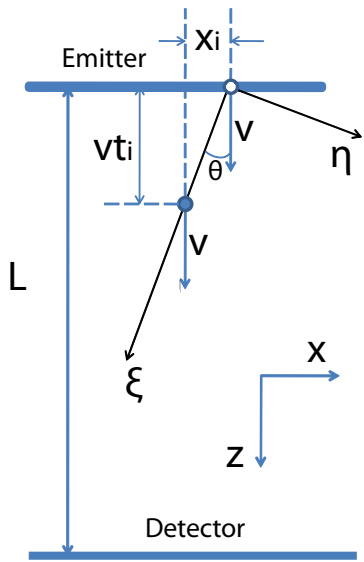}
\centering
\caption{\label{schematic} Schematic of emission of two electrons. The first (closed circle) and second (open circle) travel downward to the detector plate at velocity $\simeq v$.   Coulomb acts along the $\xi$ direction. }
\end{figure}

  The traditional method of correcting for Coulomb interactions is to employ
the Gamow correction, which assumes that the characteristic
separation of the pair of particles is much smaller than their
classical turning point, namely, that the particles are produced
well within the classically forbidden regime bounded by $r_{tp}$.
The actual rate observed in an experiment is taken to be that in
the absence of Coulomb interactions times the Gamow correction,
$|\psi_c(0)|^2$, which is the absolute value squared of the relative
Coulomb wave function at the origin,
\begin{equation}
  \psi_c (0) = \left(\frac{2 \pi \eta}{e^{2 \pi \eta}-1}
  \right)^{1/2},
\end{equation}
where the dimensionless parameter $\eta$ equals $z z' e^2/\hbar
v_{rel}$, with $v_{rel}$ the relative velocity of the two particles
of charges $ze$ and $z'e$. 
On the other hand, in a field emission source, the relative energy of a pair is  $\simeq \Delta E$,
where $\Delta E$ is the initial electron energy spread in the beam.
Thus $r_{tp} \simeq e^2/\Delta E \simeq (1.46 /\Delta E_{eV})$ nm, where $\Delta E_{eV}$ is the energy spread measured in electron volts.  Since $r_{tp}$ is generally tens of nm, 
$r_{tp} \ll r_0$, indicating that a pair of electrons
in such an experiment is
typically emitted {\it outside} the pair's classical turning point.    Coulomb effects are dominated
by classical physics rather than by a quantum Gamow correction.

\section{Classical model}

 To bring out the effect of Coulomb interactions, we model the experiment as independent emissions of electrons from a tip, followed by acceleration 
to final velocity $v$ in the beam direction (z) and energy $E_f = mv^2/2$.
We first neglect quantum
statistics, and focus on the Coulomb effects in a single pair of particles, since the 
major
contribution to the correlation function is from particles nearby in space and time,
a configuration in which we can,
to a first approximation, neglect many-body effects.   
We assume that the emission points of the pair are separated
by $x_i$ in space and $t_i$ in time.  Thus the initial spatial separation of the pair is $s_i = \left(x_i^2+(vt_i)^2\right)^{1/2}$.

\begin{figure}
\includegraphics[width=8cm]{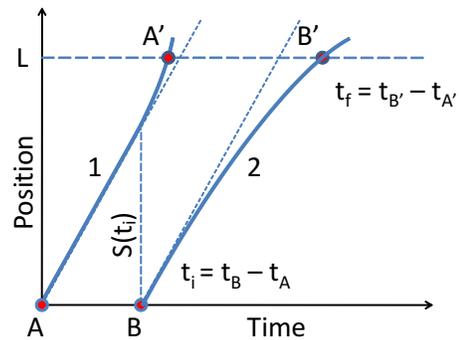}
\centering
\caption{\label{1d} Schematic picture of the effect of Coulomb repulsion on the trajectories of two electrons emitted successively from the same point in the tip. }
\end{figure}

The Coulomb repulsion between the electrons increases their relative separation in space and time, as is illustrated schematically in Fig.~\ref{1d} for two electrons emitted at the same point in the tip at times $t_A$ and then $t_B$.
The electron separation at later time is readily calculated from the conservation of energy 
of the relative motion of the two electrons, 
$
    E_{rel} =  m\dot s^2/4 + e^2/s. 
$
A lower bound on the size of the Coulomb hole can derived by neglecting the initial relative
kinetic energy of the pair; then after integration of $\dot s$, one finds that the final separation of the pair, $s_f$, is determined 
implicitly by
\begin{align}
  \label{sEq} 
  \sqrt{\frac{4e^2}{m}} \Delta t & = s_i^{3/2}
 \left\{
  \sqrt{\sigma(\sigma-1)} +\ln\left[\sqrt{\sigma} +
  \sqrt{\sigma-1}\right] \right\},
\end{align}
where $\sigma=s_f/s_i$, and $\Delta t \simeq L/v$ is the time elapsed between emission and detection of the pair. 
Approximately,
\begin{align}
\label{tf_1dEq}
   s_f=\begin{cases} s_c \left( s_c / s_i \right)^{1/2}, &
  \;\mbox{$s_i\leqslant s_c$}\\ s_i, & \;\mbox{$s_i > s_c,$}\end{cases}
\end{align}
where $ s_c  \equiv v\tau_c \equiv \left(2e^2L^2/E_f\right)^{1/3}$  
defines a critical Coulomb distance and time $\tau_c$;
numerically, $s_c \simeq 6.5\times 10^{-4}
L_{cm}^{2/3}/E_{f,KeV}^{1/3}\,{\rm cm}$.  

\begin{figure}
\begin{center}
\includegraphics[width=8cm]{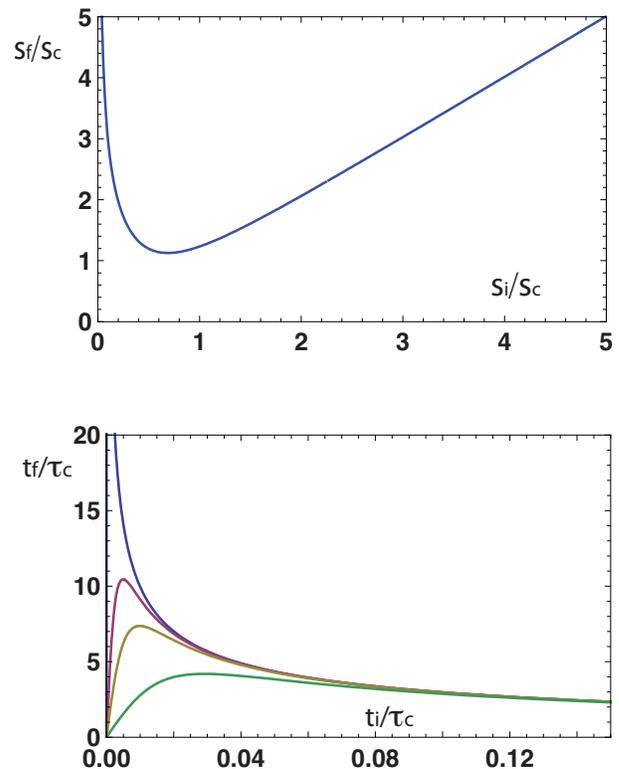}
 \caption{\label{tfti-comb} a)  Final vs. 
initial spatial separations, with Coulomb interactions included
classically, and b) final vs. initial
time separations for
different initial transverse separations $x_i$
= 0, 5, 10, and 30 nm (top to bottom).   For large $t_i/\tau_c$, the
curves all converge to that in the left panel.}emo\end{center}
\end{figure}

As seen in the left panel of Fig.~\ref{tfti-comb}, the relation between the final separation $s_f$ when the electrons reach the detector plate and $s_i$ is not monotonic, but rather is
decreasing at small $s_i$ and increasing at
large $s_i$.  For very small initial separation, the Coulomb interaction
significantly accelerates the two electrons away from each other,
making the final separation large, while for very large initial
separation, the Coulomb interaction is negligible, and the final
separation is essentially equal to that initially.
No matter how small $s_i$ is, the final spatial separation is finite, i.e., there is a {\em Coulomb hole}
in the distribution of final separations.    Coulomb forces increase the spatial separation
between a pair of electrons, $s(t)$, with time, so that
$s_f>s_i$.

Since the angle $\theta$ that the relative position vector of the electrons makes with respect to the beam axis (see 
Fig.~\ref{schematic}) is conserved in the motion, the final separations at detection are related by
$
 x_f/x_i = t_f/t_i = s_f/s_i;
$
thus the final separation in time of the two electrons is given by
$t_f = (s_f/s_i)t_i$.  At small $t_i/\tau_c$, the final $t_f$ has the structure shown in the right of Fig.~\ref{tfti-comb}, dependent on $x_i$.  However,
for $t_i/\tau_c \simge 0.1$, the
curves on the right of Fig.~\ref{tfti-comb} converge to that in the left panel of Fig.~\ref{tfti-comb}, which has a minimum at  $s_f \approx s_c$. To a first approximation, the minimum $t_f \approx s_c/v = \tau_c$ is independent of the initial spatial separation.  

\begin{figure}
\begin{center}
\includegraphics[width=8cm]{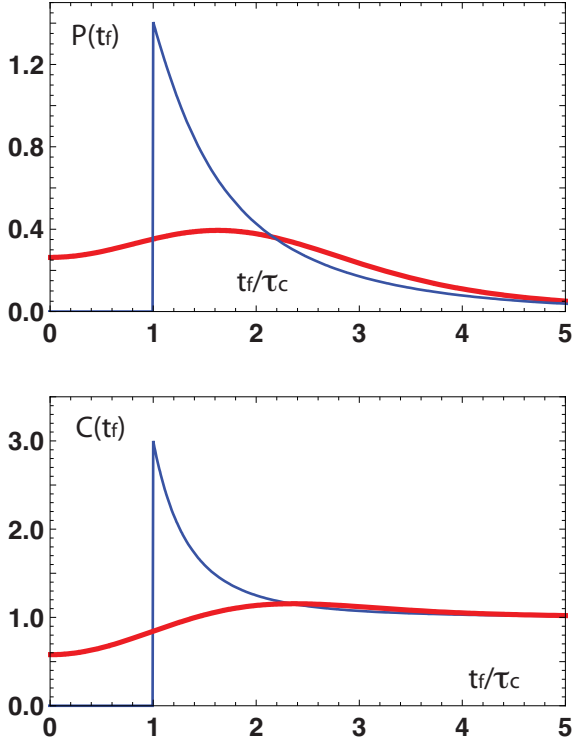}
\end{center}
\caption{\label{cp} Effects of Coulomb repulsion as seen in a) the final time distribution $P(t_f)$
vs.  $t_f$, with $\bar{t_i} = 0.2$ ns; the vertical axis is
in units of $1/\bar{t_i}$, and b) the normalized second-order final correlation function
$C(t_f)$ vs. $t_f$.  The thin lines are calculated directly from Eq.~ (\ref{tf_1dEq}); the thick lines include
finite time resoluton via Eqs.~(\ref{resolution}) and (\ref{R}), with $t_r = \tau_c$. }  
\end{figure}

  Experimentally one measures the correlation function $C(t_f)$ in terms of
the subsequent time intervals $t_f$ between detection of
particles, averaged over the distribution of initial emission intervals $t_i$.   For independent emissions, the distribution of times between
adjacent emissions from the tip is Poissonian
 \begin{equation}
  P_0(t_i) = \frac{1}{\bar{t_i}} e^{-t_i/\bar{t_i}},
\end{equation}
where $\bar{t_i}$ is the average time separation between two emissions.
 In the absence of Coulomb corrections and
quantum statistics, $C_0(t_i) = 1$.     The final $C(t_f)$ and $P(t_f)$ are given in terms of the map 
(\ref{sEq}) between $s_f$ and $s_i$; since the map is not simply one-to-one, one needs to sum over the two branches.
The resulting $P(t_f)$ and $C(t_f)$, calculated with the approximate solution (\ref{tf_1dEq}) are shown as thin lines in
Fig.~\ref{cp}.

    In experiments, the finite time resolution of the detectors would smooth out the sharp Coulomb holes in Fig.~\ref{cp}.
Measurement of an observable $\widetilde{f}(t)$
at time $t$ averages the actual $f(t')$ over $t'$,
weighted by the detector time resolution function $R(t-t')$:
\begin{equation}
\label{resolution}
   \widetilde{f}(t) = \int_{-\infty}^{+\infty} R(t-t') f(t') dt',
\end{equation}
with $f(t)=f(-t)$ for $t<0$.  In Fig.~\ref{cp}, effects of time resolution are indicated by the thick curves, where we take a Gaussian time resolution function
\begin{equation}
   R(t-t') = \frac{1}{\sqrt{2\pi} t_r}
      e^{-\left(t-t'\right)^2/2t_r^2} \,\, ,
\label{R}      
\end{equation}
with a characteristic time scale $t_r$ chosen here for illustration to equal $\tau_c$.   Typically, $t_r \gg \tau_c $.   The dip at low $t_f$ in right panel illustrates clearly how Coulomb correlations can mimic quantum correlations.

\section{HBT anti-correlations}

To assess the importance of the Coulomb hole it is necessary to compare it with the regime of suppression of the correlation function from quantum statistics.  The Pauli principle suppresses the correlation function between same spin electrons emitted from nearby points on a time scale
\beq
   t_{HBT} = \hbar/T_f,
   \label{thbt}
\eeq
where $T_f$ is the effective longitudinal temperature of the electron gas after acceleration.   Since the acceleration of the beam is essentially adiabatic, the entropy per electron is conserved, and the temperature of the gas falls with expansion.
Owing to acceleration
the density $n$ of the gas drops, since in a steady state the current $nv$ remains constant from emission through acceleration.  To estimate the final gas temperature, we note that the entropy per particle of a gas with an anisotropic temperature  depends on  $T^{1/2}T_\perp/n$, where  $T$ is the longitudinal and $T_\perp$ the transverse temperature.  Thus in anisotropic expansion, $T^{1/2}T_\perp/n$ remains constant, and for $T_\perp$ constant,
 $Tv^2$ is invariant.   Initially $v = \sqrt {T_i/m}$ and after acceleration to velocity much greater than that of the thermal motion, $v = \sqrt{ 2E_f/m}$.  Thus  the final longitudinal temperature of the beam is given by $T_f  \sqrt {2E_f/m} \simeq T_i \sqrt {T_i/m}$, so that with Eq.~(\ref{thbt}) and $T_i \simeq 2\Delta E$, we have $T_f = 2(\Delta E)^2/E_f$ (essentially the result derived heuristically in Ref.~1) and
\beq
  t_{HBT} = \frac{\hbar E_f}{2(\Delta E)^2}.
\eeq  
The ratio of the Coulomb to HBT suppressions is thus
\beq
   \frac{ t_{HBT} }{\tau_c} \simeq 2^{-5/6} \left(\frac{a_0}{L}\right)^{2/3}\frac{E_f^{11/6}{\rm Ry}^{1/6}}{(\Delta E)^2}
    = 0.9\frac{E_{f,{\rm KeV}}^{11/6}}{\Delta E_{\rm eV}^2L_{{\rm cm}}^{2/3}}.\nonumber\\
\eeq
This simple calculation indicates the importance, in an HBT measurement of correlations in arrival times, of accelerating the electrons to a large final beam energy in order to overcome Coulomb effects.   

   In addition to time anti-correlations in the beam, HBT correlations should appear in the electron spatial separations,
 analogous to the spatial correlations seen in the original Hanbury Brown--Twiss measurement of the angular diameter of the star Sirius \cite{hanbury}.  When two electrons are emitted at the same time,  but spatially separated, the correlation function is suppressed in space on a scale of order the particle wavelength divided by the angular size of the source, or
 \beq
  s_{HBT} \simeq \frac{\hbar L}{mv r_0} = 6\times 10^{-4} \frac{ L_{cm}}{E_{f,KeV}^{1/2}\,r_{10nm}}\, {\rm cm}
\eeq
where $r_{10nm}$ is the transverse size of the emission region in units of 10 nm.   Comparing with the minimal Coulomb hole, $s_c$, we find 
\beq
    \frac{s_{HBT}}{s_c} \simeq \frac{ L_{cm}^{1/3}}{E_{f,KeV}^{1/6}\,r_{10nm}}.
\eeq
To reach this ratio, the initial spatial separation must be of order $s_c$; however, a more realistic bound on $s_i$ is the transverse size of the emission tip, which leads to a considerably larger Coulomb hole, as one can infer from Fig. 1.

   In the experiment of  Kodama, Osakabe, and Tonomura \cite{KOT}  $E_f \sim$ 50-100  KeV, and $\Delta E \simeq$ 0.17 eV; with $L \sim$ 100 cm  one estimates that  $t_{HBT}/\tau_c \sim (2-6)\times 10^3$, sufficiently high that
one does not need to worry about Coulomb effects in measuring pure time-of-arrival correlations.   On the other hand,
in the opposite regime, when measuring spatial HBT correlations, one has $s_{HBT}/s_c \sim$ 0.5, indicating that Coulomb effects must be taken into account in analyzing the experiment.

   In contrast, in the lower energy experiment of Kiesel et al. \cite{kiesel}, where $E_f \sim$ 0.9 KeV, $\Delta E \simeq$ 0.13 eV, and $L \sim$ 1 cm, one has $t_{HBT}/\tau_c \sim 44$, and thus Coulomb effects while small are not entirely negligible.  For spatial correlations, however,  $s_{HBT}/s_c \sim$ 0.25. and thus Coulomb effects are dominant.
   
    In conclusion, a full analysis of the HBT experiment of Kodama, Osakabe, and Tonomura requires correcting for Coulomb effects.  The simple model presented here forms a useful and simple basis for including Coulomb interactions among the electrons.  A detailed analysis requires taking into account the distribution of initial spatial separations and velocities of the electrons in addition to the time separations,  the effects of realistic finite time resolution, as well as the effects of the quadrupole magnets which give one the freedom to adjust the angular size of the beam.   While Coulomb effects are present independent of the relative spin of the electron pair, Pauli quantum correlations occur only between same spin electrons; thus to  distinguish optimally Coulomb repulsions from quantum correlations one would ideally like to repeat the experiments with spin polarized electron beams.
    
\section*{Acknowledgments\label{sec:acknowledgements}}

  This paper is dedicated to the memory of the late Akira Tonomura.  He was at the same time a remarkable scientist and a warm friend with whom author GB spent many a happy moment, from Tokyo to Urbana and Washington to Scandinavia; his constant interest in HBT correlations in a free electron beam was a valuable source of inspiration. 

The research discussed here, based in good measure on the Ph.D. dissertation \cite{kan} of author KS, has been supported in part by the U.S. National Science Foundation over the years, most recently by NSF Grants PHY07-01611 and PHY09-69790.   GB is grateful to the Aspen Center for Physics, supported in part by NSF Grant PHY10-66293, where parts of this research were carried out.

\end{document}